\documentclass[aps,pre,showpacs,twocolumn]{revtex4-1}

\usepackage{amssymb}
\usepackage{amsmath}
\usepackage{graphicx}
\usepackage{color}
\usepackage{bm}
\usepackage[normalem]{ulem}
\begin{document}

\title{Analytical and numerical $K_u - B$ phase diagrams for cobalt nanostructures: stability region for a Bloch skyrmion}
\author{A. Riveros$^1$}
\author{N. Vidal-Silva$^1$}
\author{F. Tejo$^1$}
\author{J. Escrig$^{1,2}$}

\affiliation{$^{1}$Departamento de F\'{\i}sica, Universidad de Santiago de Chile (USACH),
Av. Ecuador 3493, 9170124 Santiago, Chile\\
$^{2}$Center for the Development of Nanoscience and Nanotechnology (CEDENNA), 9170124 Santiago, Chile
} 
\begin{abstract}
In this letter we calculate the energies corresponding to the different magnetic phases present in a ferromagnetic cylinder by means of analytical calculations. From the comparison of these energies, it is possible to construct magnetic phase diagrams as a function of the uniaxial anisotropy of the sample and the external magnetic field applied. As proof of concept, we analyzed the magnetic phase diagrams for a cobalt dot of 240 nm in diameter and 70 nm in length, with an easy axis parallel to the dot axis, and with a magnetic field applied towards or perpendicular to this axis. From these diagrams we have obtained the stability regions for a Bloch skyrmion (Sk), a vortex core (VC) and a ferromagnetic (F) configuration, which can point in any $\psi$ direction. Our results provide a pathway to engineer the formation and controllability of a skyrmion in a ferromagnetic dot to different anisotropy constants and magnetic fields.
\end{abstract}

\maketitle

\section{Introduction}

Skyrme was the first to describe the baryons as topological defects of continuous fields \cite{skyrme62}. Since then, skyrmions have been found in various systems, such as ferroelectrics \cite{NPL+15}, liquid crystals \cite{ATS+14}, magnetic materials \cite{NYT12}, among others. For example, topological magnetic skyrmions \cite{FBT+16} have been observed in several bulk \cite{MBJ+09,MNA+10,YOK+10,SYI+12} and thin film \cite{YKO+11,FSK+13,JUZ+15,MMR+16,WLK+16,WZX+16} systems, and have been proposed for potential applications in non-volatile magnetic memories \cite{RHM+13} because the spin texture topology protects the skyrmions from scattering with structural defects, allowing them to be moved by small current densities, opening a new paradigm for the manipulation of magnetization at the nanoscale \cite{JMP+10}. Besides, skyrmions exhibit emergent electromagnetic phenomena, such as topological Hall effect and the skyrmion Hall effect \cite{NT13,LLK+17}, and have been proposed as information carriers in novel magnetic sensors and spin logic devices \cite{ZEZ15}.  

Isolated skyrmions confined in cylindrical nanostructures \cite{WLK+16,MMR+16,SCR+13,RT13,DWT+13,Guslienko15,BCW+15,SA16,BVY+16,GG16} are considered to be promising for implementations in information storage and processing devices on the nanoscale \cite{FCS13,SCF+13}. In these nanostructures both the Dzyaloshinskii-Moriya interaction (DMI) and the magnetic anisotropy are required to stabilize a Neel skyrmion (NS) \cite{RT13,SCR+13,BCW+15}, where the magnetic profile has a magnetic component in the radial direction, so they cannot be seen in conventional ferromagnetic materials (Co, Ni, etc.). On the other hand, the Bloch skyrmions (BS), which do not have magnetic component in the radial direction, can be stabilized in the absence of DMI, provided there is a magnetic anisotropy \cite{Guslienko15,SA16,DWT+13}. These systems show potential for room temperature control of skyrmions \cite{MWY+15,GMB+15}. 

In this letter, we are interested in obtaining analytical expressions for the energies of different magnetic configurations (ferromagnetic pointing in any direction, vortex core and Bloch skyrmion without DMI) in a cobalt nanodot that allow us to generate magnetic phase diagrams with regions of stability for each configuration as a function of the uniaxial anisotropy and the external magnetic field. In addition, we will carry out micromagnetic simulations for some particular cases, in order to be able to support the theoretical model used.

\section{Analytical model}

We adopt a simplified description of the system, where the discrete distribution of the magnetic moments is replaced with a continuous one characterized by a slow variation of the magnetization $\vec{M} (\vec{r}) = M_0 \, \hat{m}(\vec{r})$ \cite{Aharoni96}, whose direction is given by the unitary vector $\hat{m}(\vec{r})$ while that $M_0$ corresponds to the saturation magnetization. Due to the cylindrical symmetry of the nanoparticle, it is convenient to rewrite the magnetization vector as $\hat{m} (\vec{r}) = m_r(\vec{r}) \hat{r} + m_{\phi}(\vec{r}) \hat{\phi} + m_z(\vec{r}) \hat{z}$, where $\hat{r}$, $\hat{\phi}$ and $\hat{z}$ are the unitary vectors of the cylindrical coordinates. 

We consider a cylindrical nanoparticle of radius $R$ and length $L$ which exhibits an uniaxial anisotropy whose axis of easy magnetization is parallel to the symmetry axis of the particle (chosen as the z-axis), and which is under the action of an external magnetic field $\vec{B}$ applied at an angle $\theta$ with respect to the $z$-axis, as shown in Fig. 1a. The total energy for this nanoparticle is given by
\begin{eqnarray}
\label{E_t}
E = \int_V \Bigg(-K_u m_z^2+\frac{\mu_0}{2}M_0 \, \vec{m}\cdot \vec{\nabla} U_d\nonumber \\ +A\sum_{i=x,y,z}\left( \vec{\nabla} m_i\right)^2- M_0 \, \vec{m} \cdot \vec{B} \Bigg)dV\hspace{0.3 cm} \,,
\end{eqnarray}
wherein the first, second, third and fourth term corresponds to the uniaxial anisotropy, the dipolar energy, the exchange energy and the Zeeman energy, respectively. Here $K_u$, $A$ and $\mu_0$ are the anisotropy constant, stiffness constant and magnetic permeability, respectively, while $U_d$ is the well-known magnetostatic potential defined as \cite{Aharoni96} $4\pi U_d(\vec{r})=\int G(\vec{r},\vec{r}\, ')\left(\hat{n}\cdot \vec{M}(\vec{r}\, ')- \vec{\nabla} \cdot \vec{M}(\vec{r}\, ')\right)$, with $G(\vec{r},\vec{r}\,')= |\vec{r} - \vec{r}\, '|^{-1}$ being the Green function. In the previous definition of $U_d(\vec{r})$, the first integral is over the surface of the nanoparticle, while the second is on its volume.

\begin{figure}[ht]
\centering
\hspace{-0.3cm} \includegraphics[width=1.0\linewidth]{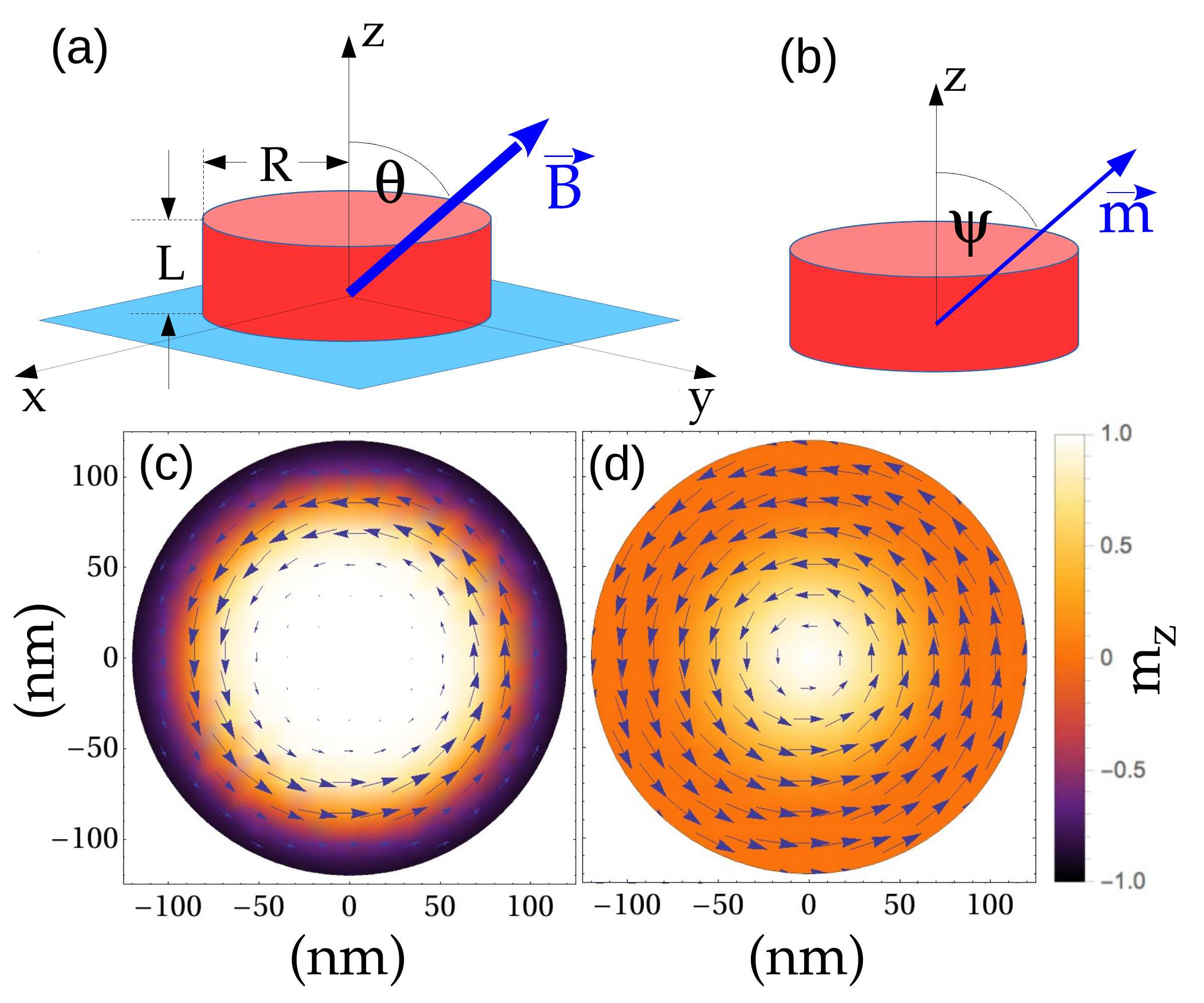} \vspace{-0.1cm}
\caption{(Color online) a) Geometrical parameters of the cylindrical nanostructure investigated under the action of a uniform magnetic field $\vec{B}$. Magnetic configurations studied: b) ferromagnetic forming an angle $\psi$ with respect to the $z$-axis, c) skyrmion ($R_s = 90$ nm, $n = 10$) and d) vortex core ($R_v= 90$ nm).}
\label{Phases_fig}
\end{figure}

We are interested in obtaining phase diagrams showing the stability regions for the following magnetic configurations: skyrmion (Sk), ferromagnetic (F) and vortex core (VC).

\subsection{Skyrmion configuration (Sk)}

For the description of a skyrmion configuration we have used the Ritz model proposed in \cite{VRE17}:
\begin{equation}
\label{mz_sky}
m_z^{(\text{Sk})}(r)=\frac{1-\left(r/R_s\right)^n}{1+\left(r/R_s\right)^n}
\end{equation}
where $R_s$ is the radius of skyrmion and $n$ is a positive even integer number, $n = 2, 4, 6, 8, \cdots$. It is important to mention that the components in the plane of the magnetization are given by $m_r = 0$ and $m_\phi= 1- m_z^2$. As an example, in Fig. 1c we show the profile of the magnetization of a skyrmion, in a nanoparticle of $R = 120$ nm, obtained from the Eq. 2 for $n = 10$ and $R_s = 90$ nm. The $m_z$ component is shown as a density color plot, while the $m_r$ component is represented by arrows.
 
\subsection{Ferromagnetic configuration (F)}

As we have considered a competition between the uniaxial anisotropy (which favors the magnetization to point along the $z$-axis) and the external magnetic field, which forms an angle $\theta$ with respect to the $z$-axis, as shown in Fig. 1a, we have used a ferromagnetic configuration whose direction is allowed to point at an angle $\psi$ with respect to the $z$-axis, as shown in Fig. 1b.
\begin{equation}
\label{m_ferro}
\hat{m}^{(\text{F})} = \cos\psi \, \hat{z} + \sin\psi \, \hat{r}
\end{equation}

\subsection{Vortex core configuration (VC)}

Finally, we have also considered a vortex core configuration, for which we have used the Ritz model previously investigated by \cite{LEA+05,MAL+10,RVL+16}
\begin{equation}
\label{mz_vc}
 m_z^{(\text{VC})}(r) = \left\lbrace \begin{matrix}
  [\, 1 - (r/R_v)^2 \, ]^2 \hspace{0.4cm}, 0 \leq r \leq R_v\\
 \\
0 \hspace{2.8cm}, \text{otherwise}
\end{matrix} \right.
\end{equation}
where $m_r = 0$ and $m_\phi =1-m_z^2$, while $R_v$ corresponds to the core size. As an example, in Fig. 1d we show the profile of the magnetization of a vortex core, in a nanoparticle of $R = 120$ nm, obtained from the Eq. 4 for $R_v = 90$ nm. 

To obtain the minimum energy configuration for a given set of geometric parameters ($R$ and $L$) and magnetic ($A$, $M_0$, $K_u$ and $\vec{B}$), we calculate the energy of each magnetic configuration, for which we replace the corresponding ansatz (Eqs. 2, 3 and 4) within Eq. 1, and we minimize with respect to $R_s$, $\psi$ and $R_v$, respectively. In the case of skyrmion, we have to choose a value of $n$, for which we have performed an analysis similar to the one performed in \cite{VRE17}, obtaining that $n = 10$ is a reliable value to correctly describe a skyrmion state.

\section{Micromagnetic simulations}

In order to validate the analytical calculations, we have investigated the minimum energy configuration of a cobalt dot of radius $R=120$ nm and length $L=70$ nm by micromagnetic simulations \cite{oommf}. In this article we have considered an out-of-plane magnetic anisotropy, which is generally obtained when the cobalt is deposited on a platinum or palladium substrate \cite{Fang2013,OMS+99,KMZ+17,MOV+17}. We have used a saturation magnetization $M_0=1.4\times 10^6$ A/m, an exchange stiffness $A=3\times 10^{-11}$ J/m and a Gilbert damping constant $\alpha$= 0.5. In addition, we use four possible initial magnetic configurations (skyrmion, vortex core, ferromagnetic out of plane and ferromagnetic in plane), which are allowed to relax as a function of $B$ and $K_u$ values, to finally compare the total energies between the different final configurations to which the system evolved. In order to obtain the results in a reasonable time, the discretization of the system was $3 \times 3 \times 5$ nm$^3$. The phase diagrams showed in Section IV were obtained using the analytical equations of Section II, nevertheless some points of these diagrams were also obtained through micromagnetic simulations.

\section{Results and phase diagrams}

The equations presented in section II are quite general and allow to investigate a magnetic dot with different geometric and magnetic parameters. As an example, and by comparing the energy curves for the different magnetic configurations, we have obtained the $K_u - B$ phase diagrams for the studied cobalt nanostructures ($M_0=1.4\times 10^6$ A/m and $A=3\times 10^{-11}$ J/m) of radius $R=120$ nm and length $L=70$ nm, in the range of $0 \leq K_u \leq 2$ MJ/m$^3$ and $0 \leq B \leq 0.5$ T for both $\theta = 90^\circ$ and $\theta = 0^\circ$, which are shown in Figs. 2 and 3, respectively. In these phase diagrams we have called $\text{F}_{0-30}$, $\text{F}_{30-60}$ and $\text{F}_{60-90}$ to the ferromagnetic configuration with $0^\circ \leq \psi < 30^\circ$, $30^\circ \leq \psi < 60^\circ$ and $60^\circ \leq \psi \leq 90^\circ$, respectively. The transition lines between two magnetic configurations were obtained analytically using steps of $0.01$ T and $0.1$ MJ/m$^3$ for $B$ and $K_u$, respectively.

\begin{figure}[ht]
\centering
\hspace{-0.3cm} \includegraphics[width=1.0\linewidth]{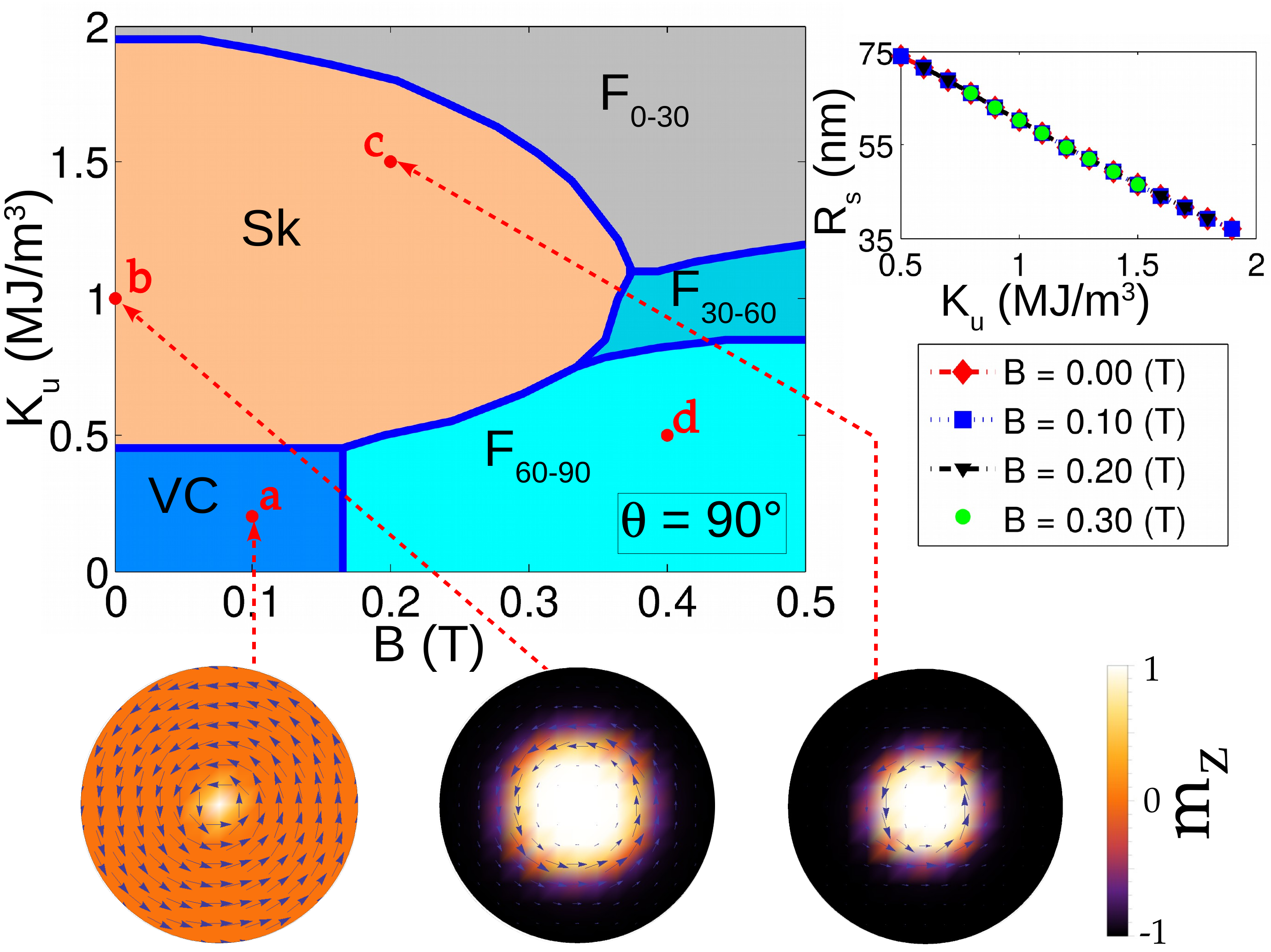} \vspace{-0.1cm}
\caption{(Color online) $K_u-B$ phase diagram for a cobalt nanostructure when the magnetic field is applied in the in-plane direction ($\theta = 90^\circ$). The four-red marked points a, b, c, and d were obtained through micromagnetic simulations, whose snapshots are shown in Fig. 4, while the three snapshots presented here correspond to the minimum energy configurations (in a, b and c) obtained from the analytical results.}
\label{PD_theta90}
\end{figure}

As can be seen from Fig. 2, when the magnetic field is in the in-plane direction, for values of $K_u < 0.5$ MJ/m$^3$ and $B < 0.17$ T the cobalt nanostructure presents a VC configuration, whereas if the uniaxial anisotropy constant increases, a Sk-phase is reported, which is stable even for magnetic fields close to $0.35$ T. On the other hand, due to the competition between uniaxial anisotropy (which favors a magnetization parallel to the $z$-axis) and the external magnetic field (which favors a magnetization perpendicular to the $z$-axis), a stable ferromagnetic phase appears, whose magnetization points in the whole range of angles, that is, $0^\circ \leq \psi \leq 90^\circ$. 

\begin{figure}[ht]
\centering
\hspace{-0.3cm} \includegraphics[width=1.0\linewidth]{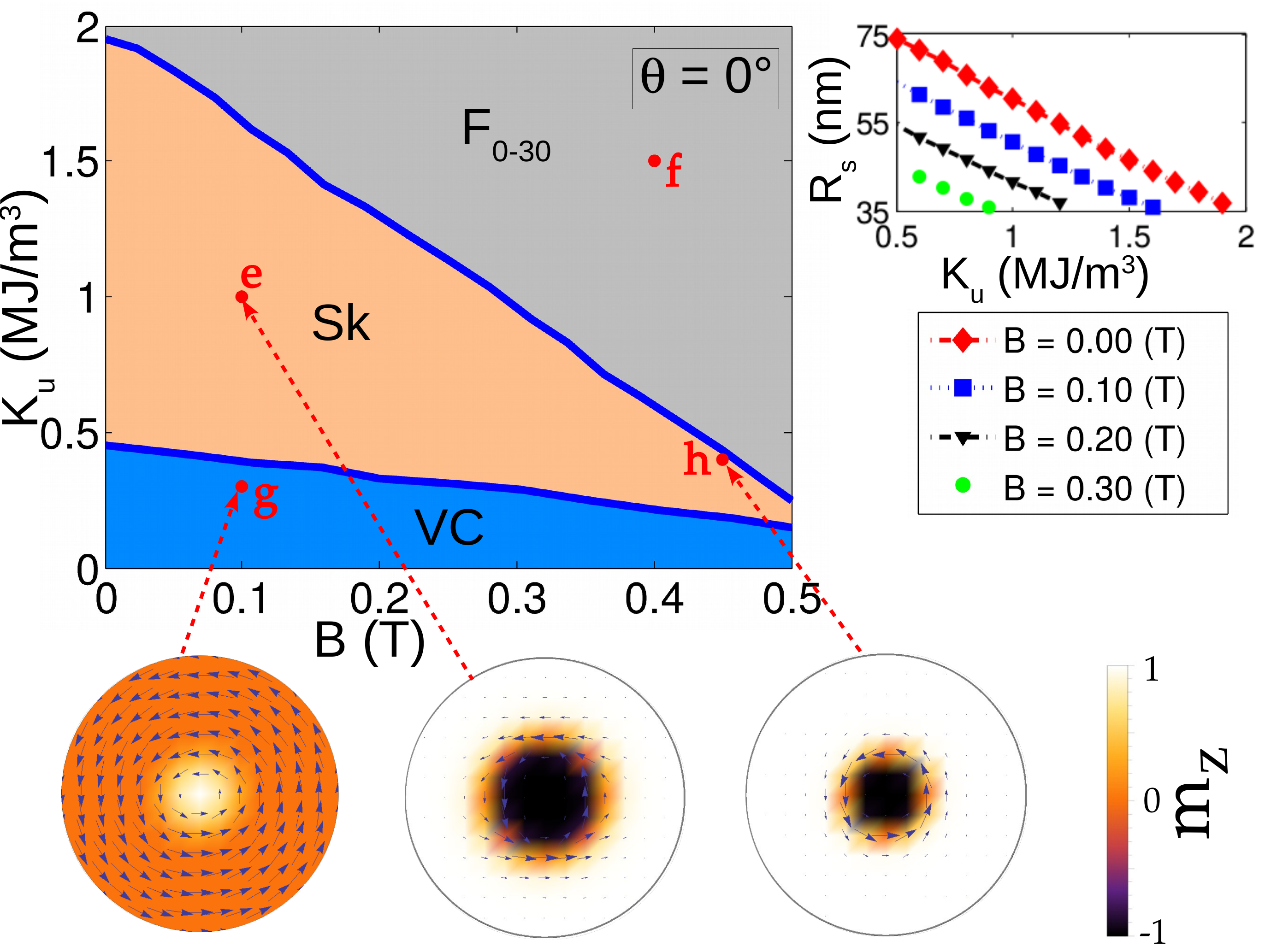} \vspace{-0.1cm}
\caption{(Color online) $K_u-B$ phase diagram for a cobalt nanostructure when the magnetic field is applied parallel to the $z$-axis ($\theta = 0^\circ$). The four red marked points e, f, g, and h were obtained through micromagnetic simulations, whose snapshots are shown in Fig. 4, while the three snapshots presented here correspond to the minimum energy configurations (in e, g and h) obtained from the analytical results.}
\label{PD_theta0}
\end{figure}

On the other hand, in Fig. 3 we analyze the situation if the magnetic field points in the same direction as the uniaxial anisotropy ($z$-axis). In this case, and although the VC phase is still present only for low $K_u$ values, it is now stable for the entire range of magnetic fields investigated. If we increase the value of $K_u$, a fairly extensive region appears where the Sk-phase is stable, covering the entire range of magnetic fields investigated. It is important to mention that when $\theta = 0^\circ$, the only surviving ferromagnetic phase is $\psi = 0^\circ$. From both figures, we can conclude that for $K_u = 0$, the Sk-phase is not stable, regardless of whether the magnetic field is applied at $\theta = 0^\circ$or at $\theta =90^\circ$.

The inset plots of Figs. 2 and 3 show the behavior of the Sk-radius as a function of the uniaxial anisotropy constant $K_u$, for different values of $B$ when $\theta = 90^\circ$ and $\theta = 0^\circ$, respectively. As can be seen $R_s$ decreases as $K_u$ increases, nevertheless for $\theta = 90^\circ$, $R_s$ does not depend on the intensity of the magnetic field, instead for $\theta = 0^\circ$, where the $Rs$-curves change for different $B$-values. Indeed $R_s$ decreases as $B$ increases, due to the core of the Sk-magnetization ($r < R_s$) points in the opposite direction of the magnetic field, in order to minimize its energy. We have compared the Sk-energy curves for cores pointing in both directions ($+z$ and $-z$), founding that in the whole Sk-phase of Fig. 3, the Sk-energy curve with core in the opposite direction of $\vec{B}$ is always below of the corresponding Sk-energy curve with core in the same direction of the field, while for $\theta = 90^\circ$, both Sk-energy curves have the same values.

Importantly, from the developed micromagnetic simulations, and for the geometric and magnetic parameters investigated in this paper, we have only obtained the theoretically proposed magnetic configurations, and we have not observed complex phases such as the helical magnetic phase. In addition, it is worth mentioning that the numerical results (showed in Fig. 4) for the marked red points in Figs. 2 and 3 have an extraordinary agreement with the analytical phase diagrams. Figure 4a shows that the vortex core is slightly offset from the center of the cylinder, which could slightly reduce the energy of this configuration. This breaking of azimuthal symmetry is out of the focus of this article.

\begin{figure}[ht]
\centering
\hspace{-0.3cm} \includegraphics[width=1.0\linewidth]{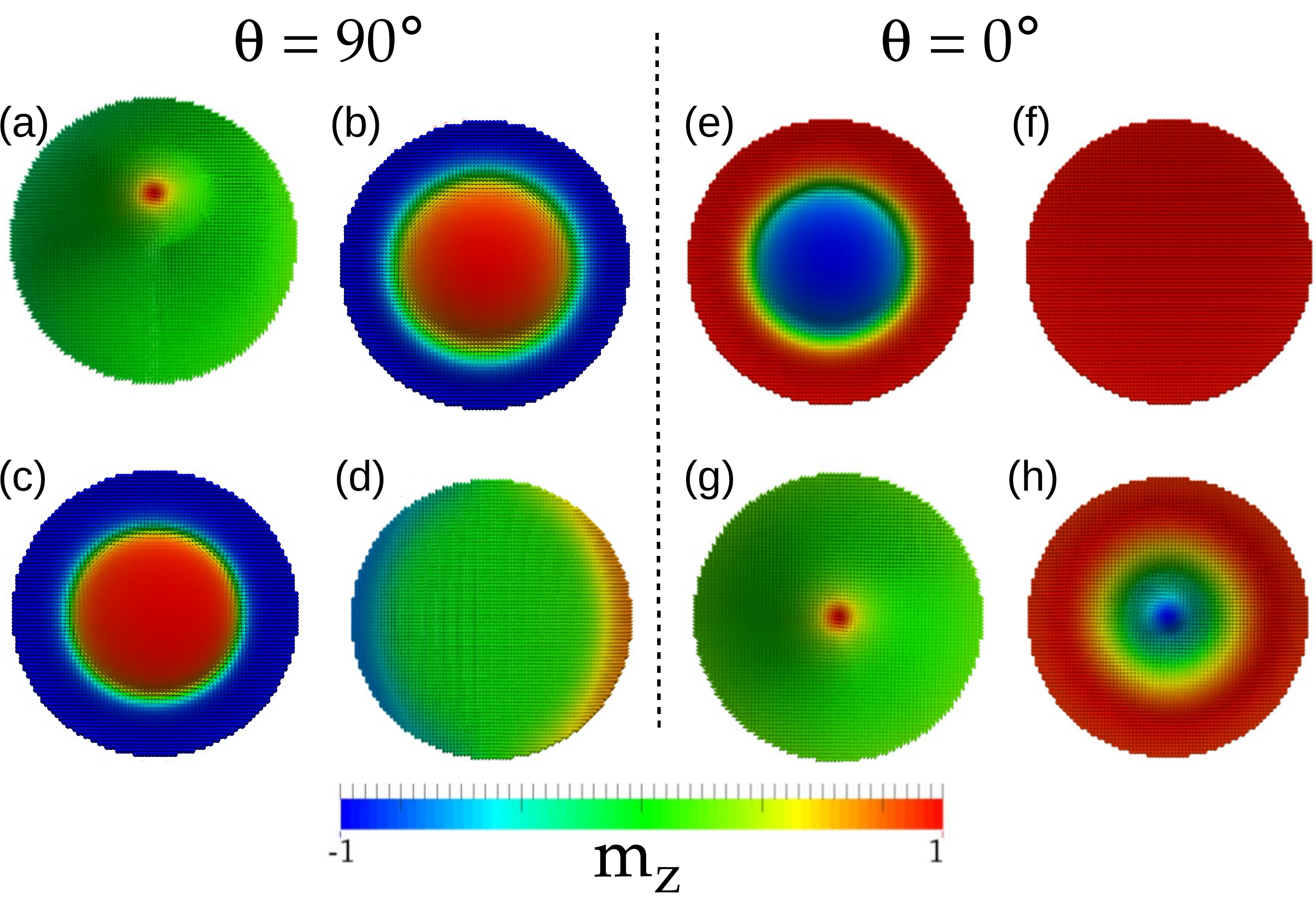} \vspace{-0.1cm}
\caption{(Color online) Snapshots of the stable magnetization states obtained through micromagnetic simulations for the eight marked points of Figs. 2 and 3: a, b, c, and d ($\theta = 90^\circ$) and e, f, g, and h ($\theta = 0^\circ$). It can be seen that both analytical and simulation results are in perfect agreement.}
\label{snapshot_sim}
\end{figure}

\section*{Conclusions}

In conclusion, by the analytical calculation of the energies corresponding to the different magnetic phases present in a ferromagnetic dot, we were able to prepare the $K_u-B$ magnetic phase diagrams for a cobalt dot of radius $R=120$ nm and length $L=70$ nm presenting the stability region for a Bloch skyrmion (Sk), a vortex core (VC) and a ferromagnetic (F) configuration. In general, and regardless of the angle at which the external magnetic field is applied, a cobalt dot will exhibit a Bloch skymion for $K_u > 0.5$ MJ/m$^3$ and low magnetic field values. It is important to mention that the radius of the skyrmion decreases with an increase in the uniaxial anisotropy constant, and as the magnetic field intensity increases (for a magnetic field pointing towards the $z$-axis), for which case we found that the skyrmion core points in the opposite direction of the magnetic field. This preferred core direction disappears when the magnetic field points in the direction perpendicular to the symmetry axis, and the skyrmion radius becomes independent of the intensity of the magnetic field. In addition, the results have been validated by micromagnetic simulations, which exhibit an excellent agreement with the analytical results. Finally, these analytical results which allow to obtain magnetic phase diagrams with the stability region of Bloch skyrmions, will be key for the design of future devices based on the manipulation of the magnetic skyrmions.

\section*{Acknowledgements}
We thank V. Salinas-Barrera for his insightful commments. This work was supported by Fondecyt Grant 1150952, DICYT Grant 041731EM-POSTDOC from VRIDEI-USACH, Financiamiento Basal para Centros Cient\'{\i}ficos y Tecnol\'{o}gicos de Excelencia FB0807, and Conicyt-PCHA/Doctorado Nacional/2014.

\end{document}